\documentclass[aps,prl,twocolumn,superscriptaddress]{revtex4}

\usepackage{graphicx}
\usepackage[squaren,thinspace,textstyle]{SIunits}

\usepackage{color}

\begin{document}

\title{Many-body correlations of electrostatically trapped dipolar excitons}

\author{}

\affiliation{}

\author{G. J. Schinner}
\affiliation{Center for NanoScience and Fakult\"at f\"ur Physik,
Ludwig-Maximilians-Universit\"at,
Geschwister-Scholl-Platz 1, 80539 M\"unchen, Germany}

\author{J. Repp}
\affiliation{Center for NanoScience and Fakult\"at f\"ur Physik,
Ludwig-Maximilians-Universit\"at,
Geschwister-Scholl-Platz 1, 80539 M\"unchen, Germany}

\author{E. Schubert}
\affiliation{Center for NanoScience and Fakult\"at f\"ur Physik,
Ludwig-Maximilians-Universit\"at,
Geschwister-Scholl-Platz 1, 80539, Germany}

\author{A. K. Rai}
\affiliation{Angewandte Festk\"orperphysik, Ruhr-Universit\"at Bochum, Universit\"atsstra{\ss}e 150, 44780 Bochum, Germany}

\author{D. Reuter}
\affiliation{Angewandte Festk\"orperphysik, Ruhr-Universit\"at Bochum, Universit\"atsstra{\ss}e 150, 44780 Bochum, Germany}

\author{A. D. Wieck}
\affiliation{Angewandte Festk\"orperphysik, Ruhr-Universit\"at Bochum, Universit\"atsstra{\ss}e 150, 44780 Bochum, Germany}

\author{A. O. Govorov}
\affiliation{Center for NanoScience and Fakult\"at f\"ur Physik,
Ludwig-Maximilians-Universit\"at,
Geschwister-Scholl-Platz 1, 80539 M\"unchen, Germany}
\affiliation{Department of Physics and Astronomy, Ohio University, Athens, Ohio 45701}

\author{ A. W. Holleitner}
\affiliation{Walter Schottky Institut and Physik-Department, Am Coulombwall 4a, Technische Universit\"at M\"unchen, D-85748 Garching, Germany}

\author{J. P. Kotthaus}
\affiliation{Center for NanoScience and Fakult\"at f\"ur Physik,
Ludwig-Maximilians-Universit\"at,
Geschwister-Scholl-Platz 1, 80539 M\"unchen, Germany}

\begin{abstract}
We study the photoluminescence (PL) of a two-dimensional liquid of oriented dipolar excitons in $\mathrm{In}_{x}\mathrm{Ga}_{1-x}\mathrm{As}$ coupled double quantum wells confined to a microtrap. Generating excitons outside the trap and transferring them at lattice temperatures down to $T=240$\,mK into the trap we create cold quasi-equilibrium bosonic ensembles of some 1000 excitons with thermal de Broglie wavelengths exceeding the excitonic separation. With decreasing temperature and increasing density  $n\lesssim5\times10^{10}\frac{1}{\mathrm{cm}^{2}}$ we find an increasingly asymmetric PL lineshape with a sharpening blue edge and a broad red tail which we interpret to reflect correlated behavior mediated by dipolar interactions. From the PL intensity $I(E)$ below the PL maximum at $E_{0}$ we extract at $T<$\,5\,K a distinct power law $I(E) \sim (E_{0}-E)^{-|\alpha|}$ with -$|\alpha| \approx$ -0.8  in the range $E_{0}-E$ of 1.5-4\,meV, comparable to the dipolar interaction energy.
\end{abstract}

\maketitle

Weakly interacting bosons confined in an external potential and cooled to very low temperatures such that the thermal de Broglie wavelength becomes comparable to the inter-particle distance can form a Bose-Einstein condensate (BEC). In this new state of matter, a large fraction of the bosons condense into the energetically lowest quantum state of the external potential, and form a correlated state with a macroscopic wavefunction. BEC has been observed in different systems such as ultra cold diluted atomic gases \cite{1995Ketterle, 1995Anderson} or superfluids \cite{1999Leggett}. More recently, condensation of non-equilibrium quasi-particles in solids, namely cavity exciton polaritons \cite{2006Kasprzak, 2007Balili, 2010DengRevModPhys} and magnons \cite{2006Demokritov}, have been reported as other examples of BEC. In an effort to realize BEC of excitons in quasi-equilibrium, predicted already in the 1960s \cite{1962Blatt, 1968KeldyshKozlov}, coupled double quantum wells containing two-dimensional systems of long-living and spatially indirect excitons (IX) have been intensively explored for more than a decade [see, e.\,g.  \cite{1995Butov, 2002Butov_N, 2002Timofeev_JETP, 2005Rapaport, 2005Voros, 2006Voros, 2006Yang, 2008SternMott, 2009Vogele, 2009HighNanoL, 2011Alloing, 2011Schinner}], but fully convincing signatures of a BEC ground state are still missing. Such IX consist of an electron and a hole, spatially separated in the adjacent wells of coupled double quantum well (CDQW) heterostructures and bound by their Coulomb attraction. With their oriented dipolar nature they form a complex and rather strongly interacting bosonic systems with internal structure and not yet fully understood ground state properties. Only recently the influence of dipolar interactions on the ground state properties of such excitonic ensembles has been more intensely theoretically investigated \cite{2009Laikhtman}.

\begin{figure}[h]
\includegraphics[width=8.6cm]{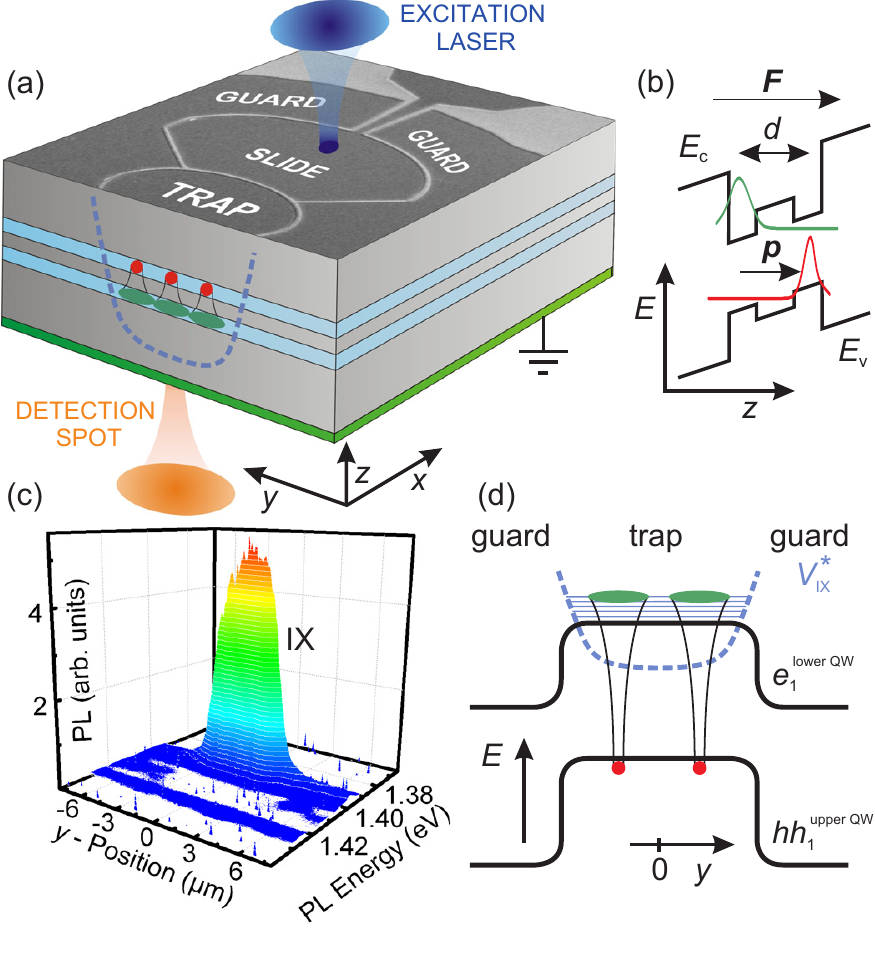}
\caption{\label{fig1} (Color online) (a) Sketch of the trapping device and detection scheme, incorporating the projection of a scanning electron micrograph of the trap (T) (with diameter of 6\,$\mu$m), slide (S), and two guard gates (G). Voltages $V_{\mathrm{T}}$, $V_{\mathrm{S}}$ and $V_{\mathrm{G}}$ are applied with respect to the back contact (green). (b) Schematic CDQW band diagram. $E_{\mathrm{c}}$ and $E_{\mathrm{v}}$ denote conduction and valence band. Electron and hole wavefunction of ground states $e_{1}^{\mathrm{lower QW}}$ and $hh_{1}^{\mathrm{upper QW}}$ are indicated in green and red, respectively. (c) Pseudo-3D picture of excitonic PL from the trap with spatial and energetic resolution [$V_{\mathrm{G}}$ = 0.70\,V, $V_{\mathrm{T}}$ = 0.35\,V, and $V_{\mathrm{S}}$ = 0.45\,V, $T_{\mathrm{Lattice}}$ = 4\,K, $P_{\mathrm{Laser}}$ = 8.0\,$\mu$W, $E_{\mathrm{Laser}}$ = 1.494\,eV]. (d) Corresponding ground state energies of electrons in the lower QW ($e_{1}^{\mathrm{lower QW}}$) and heavy holes in the upper QW ($hh_{1}^{\mathrm{upper QW}}$) versus $y$-position across the trap. The IX trapping potential $V_{\mathrm{IX}}^{*}$ is shown containing a hybridized dipolar liquid.}
\end{figure}

To achieve a quantum phase transition, an efficient trapping of suitable IX densities thermalized to cryogenic temperatures is needed \cite{2000Petrov, 2006Voros}. Going beyond previous experiments, we made special efforts to avoid a thermal imbalance between the temperature of the trapped IX and the lattice. In choosing the InGaAs material systems and resonantly exciting direct excitons in the CDQW at an energy below the band gap of the GaAs substrate we assure that most absorption of laser light occurs only in CDQW under the tight focus of the exciting laser. In addition, our special micron scale trap design sketched in Fig.\ \ref{fig1}(a) serves to generate IX on the slide gate several micrometer away from the trap and collect only precooled and neutral bosonic IX under the trap gate. The electrostatic trap uses the in-plane variation of the quantum confined Stark effect (QCSE) to spatially separate the generation of IX from the IX recombination in the trap. This allows to cool the device in a $^{3}$He refrigerator down to lattice temperatures of 240\,mK and to study the photoluminescence (PL) of typically several 1000 trapped IX with high spatial, energetic, and temporal resolution. In the corresponding IX density regime and with decreasing temperature at which the thermal de Broglie wavelength becomes larger than the excitonic spacing, we observe the integrated intensity of the IX line to increase, while the linewidth narrows and a characteristic lineshape with a sharp blue edge develops. Since this happens at temperatures and IX densities, at which the kinetic energy of the dipolar excitons falls well below their repulsive interdipolar energies of typically 2\,meV, we interpret the characteristic change in the PL lineshape and intensity as a signature of a correlated liquid of bosonic IX as theoretically discussed in Ref. \cite{2009Laikhtman}. Increasing the IX density at lowest temperature by increasing the power of the incident laser we also observe an initial decrease of the PL linewidth combined with an increase of the PL intensity and lineshape asymmetry. Analyzing the PL lineshape $I(E)$ in the energy range $E_{0}-E<$1.5-4\,meV below the PL maximum at $E_{0}$ we find a characteristic temperature-independent power law $I(E) \sim (E_{0}-E)^{-|\alpha|}$, discernable at temperatures $T<5$\,K and moderate densities and indicating an edge-like singularity at $E_{0}$.

In our device two 7\,nm thick $\mathrm{In}_{0.11}\mathrm{Ga}_{0.89}\mathrm{As}$ quantum wells (QW), separated by a 10\,nm thick GaAs barrier, are embedded between a n-doped GaAs back contact and 6\,nm thick semi-transparent titanium gate electrodes defined by e-beam lithography [Fig.\ \ref{fig1}(a)] \cite{2011Schinner}. Voltages applied to the gates define an electric field $\mathbf{F}$ perpendicular to the QW plane, and cause a red shift of the IX energy of $\Delta E_{\mathrm{IX}}=-\mathbf{pF}$, with $\mid\mathbf{p}\mid=ed$ and $d$ the electron-hole distance [Fig.\ \ref{fig1}(b)]. In-plane variation of $\mathbf{F}$ thus guides IX from the generation spot, where IX are generated via resonantly pumped direct intra well excitons, into the tunable trap \cite{SupplementaryPRL}. Employing a confocal microscope with two objectives we can separate exciton generation and PL detection, measured in transmission, at different locations \cite{SupplementaryPRL}. An image of the PL intensity in Fig.\ \ref{fig1}(c) demonstrates the trapping efficiency. As sketched in Fig.\ \ref{fig1}(d), the trapping potential $V^{*}_{\mathrm{IX}}$ is attractive for neutral dipolar IX but it repels unbound electrons. The low mobility of holes at low temperatures combined with them being trapped by the in-plane disorder potential let us assume that holes are essentially localized. In contrast, the wavefunctions of lighter electrons, confined in the lower QW and bound to holes localized in the upper QW as IX, might overlap and hybridize as their effective Bohr radius $r_{\mathrm{e}} \approx$ 20\,nm is comparable to half the excitonic separation $d_{\mathrm{IX}}$ ($d_{\mathrm{IX}}/2 \approx 28$\,nm at $n_{\mathrm{IX}} = n_{\mathrm{e}} = 4\times10^{10}\frac{1}{\mathrm{cm}^{2}}$) \cite{SupplementaryPRL} and the QW separation $d \approx$ 17\,nm, as indicated in Fig.\ \ref{fig1}(d). Caused by the repulsive dipolar interaction of the bound IX in the hybridized system one may expect correlated many-body behavior, similar to a Fermi edge singularity \cite{1987Skolnick}.

\begin{figure}[b]
\includegraphics[width=8.6cm]{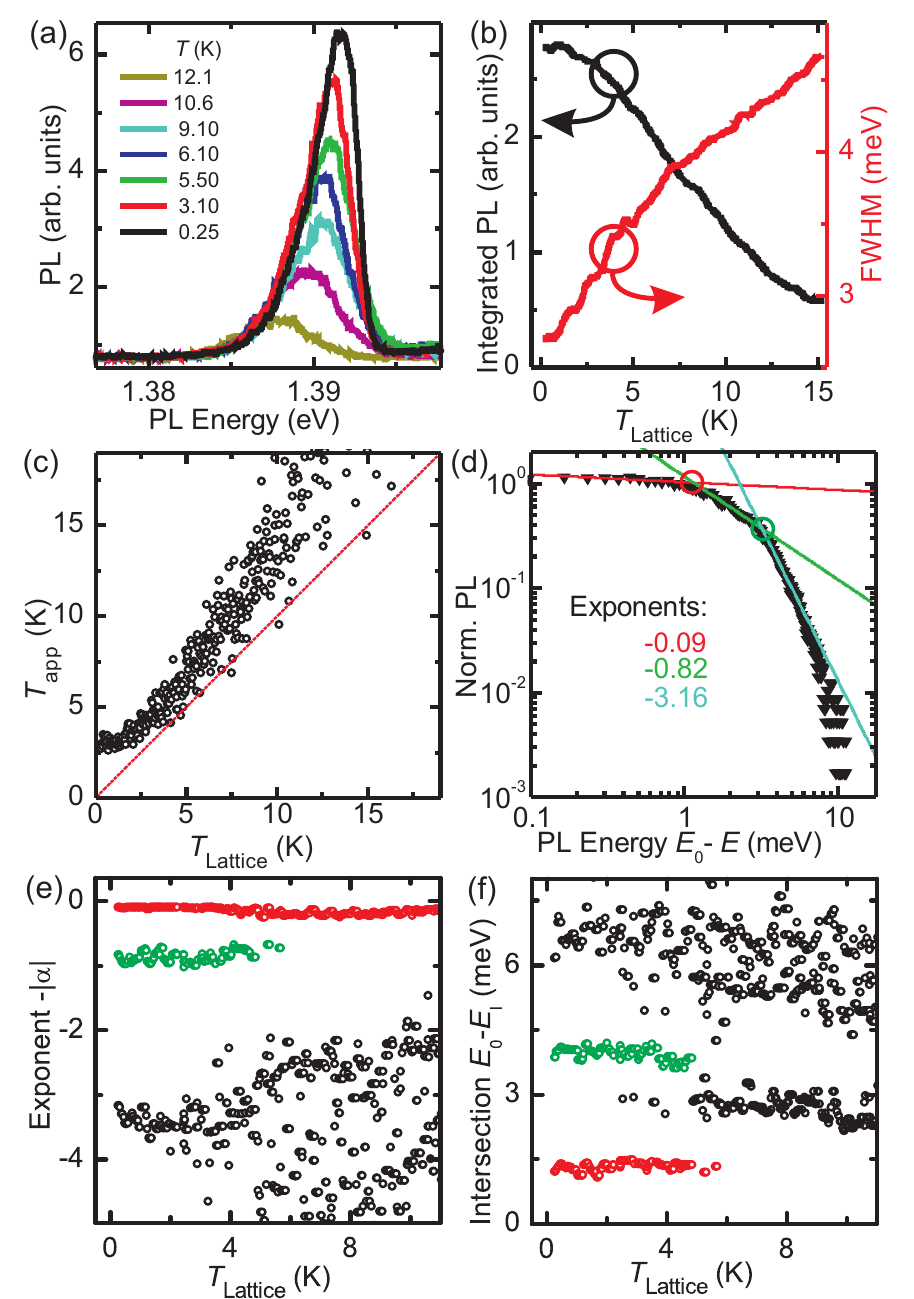}
\caption{\label{fig2} (Color online) (a) Energy-resolved PL spectra, for different lattice temperatures, detected from the center of the trap [$V_{\mathrm{G}}$ = 0.70\,V, $V_{\mathrm{T}}$ = 0.32\,V, and $V_{\mathrm{S}}$ = 0.35\,V, $P_{\mathrm{Laser}}$ = 2.7\,$\mu$W, $E_{\mathrm{Laser}}$ = 1.4289\,eV]. (b) Corresponding integrated PL intensity (left axis) and full linewidth at half maximum (right axis). (c) Exciton gas temperature $T_{\mathrm{app}}$ extracted from the blue tail of the PL line [$P_{\mathrm{Laser}}$ = 1.4\,$\mu$W]. (d) Red tail of PL normalized logarithmically plotted as a function of $E_{\mathrm{0}}-E$, with $E_{\mathrm{0}}$ the energy of the PL maximum [$T_{\mathrm{Lattice}}$ = 250\,mK]. Color-coded (e) corresponding exponents and (f) intersection energies $E_{\mathrm{0}}-E_{\mathrm{I}}$ at which the power laws in (d) change versus $T_{\mathrm{Lattice}}$.}
\end{figure}

The trap collects a cold IX ensemble of varying density and a typical lifetime of about 200\,ns caused by the reduced overlap of the electron and hole wavefunctions forming the IX.
To explore the exciton thermodynamics, we analyze the temperature dependence of the IX PL collected from the center of the trap. With decreasing temperature an increasingly asymmetric PL lineshape is observed as exemplarily shown in Fig.\ \ref{fig2}(a).
As displayed in Fig.\ \ref{fig2}(b), we also observe a strong increase of the integrated PL intensity accompanied by a narrowing of the PL linewidth.
From these experimental observations, we deduce that many-body correlations start to become important with decreasing temperature, finally resulting in an edge-like singularity.
To achieve a deeper insight into the underlying  exciton dynamics, we analyze the high (blue) and low (red) energy sides of the PL line separately. From the blue side of the PL lineshape we extract an
apparent exciton gas temperature $T_{\mathrm{app}}$ inside the trap \cite{SupplementaryPRL}. To this end, we assume only temperature-induced broadening and neglect all other broadening, e.g., by disorder [Fig.\ \ref{fig2}(c)].
Using local resonant generation of direct excitons in the CDQW we minimize radiative heating \cite{SupplementaryPRL}. Since the thermal equilibration via phonons happens on time scales below 1\,ns, much shorter than the IX lifetime, we estimate the real IX temperature in the trap to be much closer to $T_{\mathrm{Lattice}}$ than to $T_{\mathrm{app}}$. The saturation of $T_{\mathrm{app}} \simeq 3$\,K at low temperatures, corresponding to energy of $\sim0.3$\,meV, is likely to be caused by inhomogeneous line broadening and temporal fluctuations.

Many-body processes can be observed in characteristic PL lineshapes \cite{1987Skolnick, 2010NatPhKleemans}.
From recently published theoretical \cite{2011Tureci} and experimental works \cite{2011Latta} on lineshapes in quantum structures, one can conclude that different many-body phenomena with distinct energy scales are reflected by different power laws in the lineshape of discrete optical transitions at energy $E$. Such power laws correspond to different temporal dynamics, becoming faster with increasing $E_{\mathrm{0}}-E$ \cite{2011Tureci}, where $E_{\mathrm{0}}$ denotes the energy of the PL maximum. In previous PL experiments, the red tail of the spectra was attributed to the random disorder potential \cite{1991Kash, 2008Duarte_BGR} or the interplay of disorder potential and BEC \cite{2002Timofeev_JETP}. As in Refs. \cite{2011Tureci, 2011Latta}, we analyze the red tail of the PL lineshape by plotting the PL intensity logarithmically versus the logarithm of the PL energy in a $\log(I) $ vs. $\log(E_{\mathrm{0}}-E)$ diagram [Fig.\ \ref{fig2}(d)]. In the $\log$-$\log$ display it is possible to identify power laws $I(E) \sim (E_{0}-E)^{-|\alpha|}$ with different exponents $\alpha$ separated by kinks  at which different power laws intersect \cite{SupplementaryPRL}. In Fig.\ \ref{fig2}(d), and (e), we identify three different exponents associated with characteristic energy intervals [Fig.\ \ref{fig2}(f)] related to specific interactions \cite{SupplementaryPRL}. The appearance of a few exponents in an optical correlation function seems to be typical for many-body correlated systems \cite{2011Tureci}.
The most interesting exponent is the second exponent showing values of about -0.8.
We relate this to a correlated low temperature many body phase of a two-dimensional dipolar liquid of parallel oriented excitons as theoretically predicted in Ref.\ \cite{2009Laikhtman}. This power law rather abruptly disappears at $T_{\mathrm{Lattice}} \gtrsim$ 5\,K [Fig.\ \ref{fig2}(e)]. As predicted in Ref.\ \cite{2009Laikhtman} a correlated dipolar liquid is expected only at sufficiently low temperature and high enough trapped IX densities and should exhibits a rather gradual transition.

\begin{figure}[t]
\includegraphics[width=8.6cm]{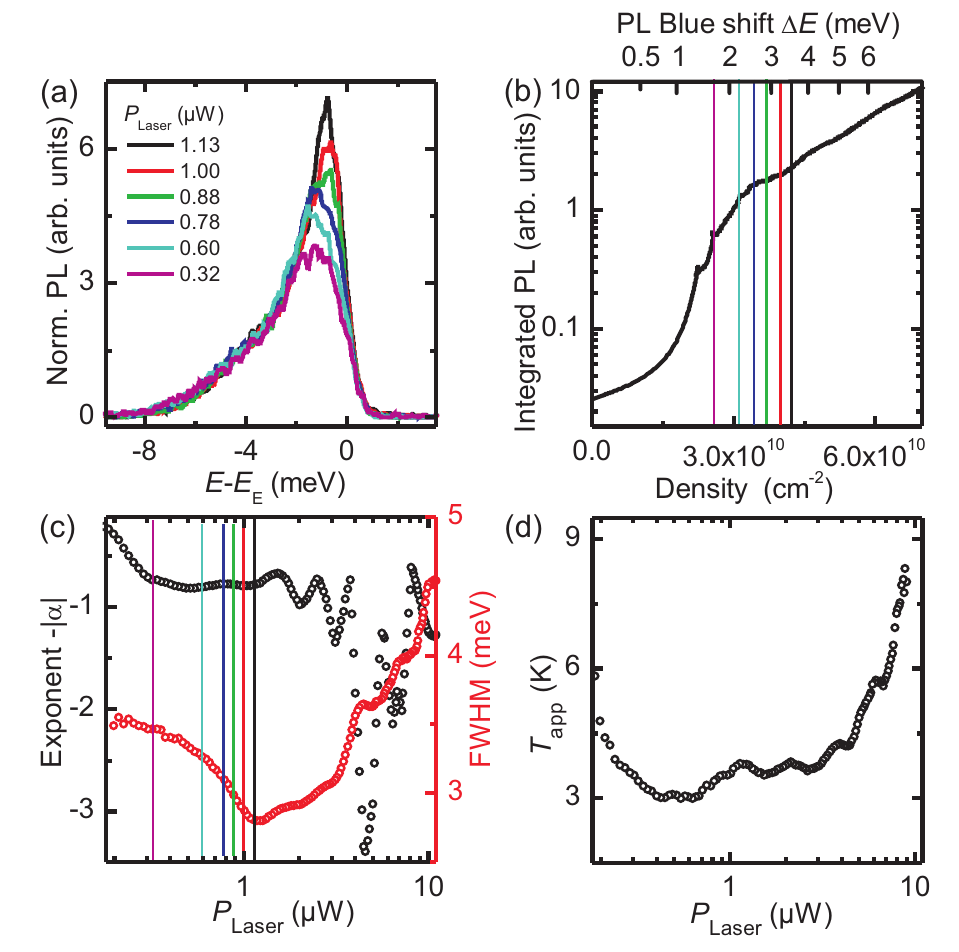}
\caption{\label{fig3} (Color online) (a) PL spectra, for different laser powers, detected from center of the trap. PL intensity is normalized to $P_{\mathrm{Laser}}$ and the spectra are energetically shifted such that the Fermi-edge-like steepest slope in the PL vs. $E$ spectra is at $E_{\mathrm{E}} \approx 0$ (original spectra see \cite{SupplementaryPRL}). (b) Integrated PL intensity plotted as a function of blue shift $\Delta E$ and corresponding density. Vertical lines in (b) and (c) are color-coded to reflect the different laser powers in (a). (c) Second exponent $\alpha$ (left axis) and linewidth (right axis) as a function of $P_{\mathrm{Laser}}$. (d) An upper limit is shown for $T_{\mathrm{app}}$ [$V_{\mathrm{G}}$ = 0.70\,V, $V_{\mathrm{T}}$ = 0.32\,V, $V_{\mathrm{S}}$ = 0.35\,V, $T_{\mathrm{Lattice}}$ = 243\,mK, $E_{\mathrm{Laser}}$ = 1.4289\,eV].}
\end{figure}

We further investigate the PL spectra as a function of the exciton density, varied by changing the exciting laser power. In Fig.\ \ref{fig3}(a), the PL intensity, normalized to the laser power, is plotted versus energy relative to the blue edge energy $E_{\mathrm{E}}$ at which the PL intensity drops to half of the maximum value (original spectra see \cite{SupplementaryPRL}). Around the PL maximum, the intensity rises more steeply than laser power $P_{\mathrm{Laser}}$. Using the blue shift $\Delta E$ of the PL maximum with increasing laser power as a measure of the exciton density, we obtain the density dependence of the integrated PL intensity shown in Fig.\ \ref{fig3}(b) \cite{SupplementaryPRL}. We point out that only a part of the nonlinear increasing PL intensity can be explained by a decreasing IX lifetime \cite{2011Schinner}.
For $P_{\mathrm{Laser}}<$ 1.2\,$\mu$W, we observe a decreasing linewidth with rising exciton density [Fig.\ \ref{fig3}(c)]. A related observation in quantum wells was recently associated with screening of the disorder potential by repulsive IX-IX interaction \cite{2011Alloing}. The density dependent decrease of the linewidth is accompanied by finding the exponent -$|\alpha| \approx$ -0.8 in the $\log$-$\log$ plot of the red tail [Fig.\ \ref{fig3}(c)]. In the same density regime we find at low temperatures the exponent -$|\alpha| \approx$ -0.8 in the temperature-dependent spectra [Fig.\ \ref{fig2}(e)] which supports the our that this power law is associated with a correlated dipolar liquid.
With increasing exciton density the dipolar interaction energies of the IX sum up to the observed blue shift $\Delta E$ \cite{SupplementaryPRL}.
The lineshape asymmetry thus can be interpreted as a broadened sum of individual excitonic transitions.
In related studies on electrostatic traps fabricated on the same InGaAs heterostructure with an effective trap diameter of down to about 100\,nm and thus a trap area of roughly 1000 times smaller than those considered here, we observe discrete individual lines of linewidth below 0.5\,meV which depend on laser power as characteristic for single excitons, biexcitons and triexcitons, respectively. Further increasing the trap population, these merge at low temperatures into an asymmetric lineshape  thus lending additional support to the interpretation given here in terms of correlated behavior.
At high laser powers, the extracted exponent $\alpha$ starts to oscillate due to an interference substructure overlaying the PL lineshape [Fig.\ \ref{fig3}(c), for details see \cite{SupplementaryPRL}] and we see an expected homogeneous broadening \cite{2009HighNanoL} of the PL line caused by exciton-exciton scattering with increasing exciton density. From the blue side of the PL lineshape we extract a rising upper limit for the apparent exciton gas temperature $T_{\mathrm{app}}$ [Fig.\ \ref{fig3}(d)].

In conclusion, at temperatures and densities at which the thermal de Broglie wavelength ($\approx 75$\,nm at 2\,K) is larger than the interexcitonic distance $d_{\mathrm{IX}} \approx 57$\,nm \cite{SupplementaryPRL} at $n=4\times10^{10}\frac{1}{\mathrm{cm}^{2}}$ we observe a characteristic temperature and density-dependent transition to a specific quantum state of the IX ensemble. In this regime, $d_{\mathrm{IX}}$ is comparable to twice the effective excitonic Bohr radius $r_{\mathrm{e}} \approx$ 20\,nm \cite{SupplementaryPRL} as well as the QW separation ($d\approx$ 17\,nm). From the appearance of the strongly increasing integrated PL intensity as well as a PL linewidth decreasing with increasing density, we conclude that this state reflects a correlated many-body state mediated by dipole-dipole interactions. In the analysis of the red tail of the PL lineshape we observe at $T_{\mathrm{Lattice}} \lesssim$ 5\,K a characteristic power law dependence of the PL intensity reflected in exponent -$|\alpha| \approx$ -0.8, associated with an energy between 1.5\,meV and 4\, meV. This we interpret as a generic fingerprint of dipolar many-body correlations.
A theoretical understanding of this power law cannot be given at present but this very interesting observation in our opinion should motivate further theoretical and experimental studies in the field.

\begin{acknowledgments}
We thank M.\,P.\,Stallhofer, D.\,Taubert, S.\,Ludwig, K.\,Kowalik-Seidl, H.\,Stolz and H.\,Tureci for stimulating discussions and helpful comments. Financial support by the DFG under Project No.\ Ko 416/17, the SPP1285 as well as the German Excellence Initiative via the Nanosystems Initiative Munich (NIM), LMUexcellent and BMBF QuaHL-Rep (01 BQ 1035) is gratefully acknowledged.
\end{acknowledgments}

\renewcommand{\thefigure}{S\arabic{figure}}
\setcounter{figure}{0}

\section{Supplemental Material: Many-body correlations of electrostatically trapped indirect excitons }

\section{Cryogenic setup}\vspace{2mm}

\begin{figure}[htb]
\vspace{14mm}
\begin{center}
\includegraphics[width=8.6cm]{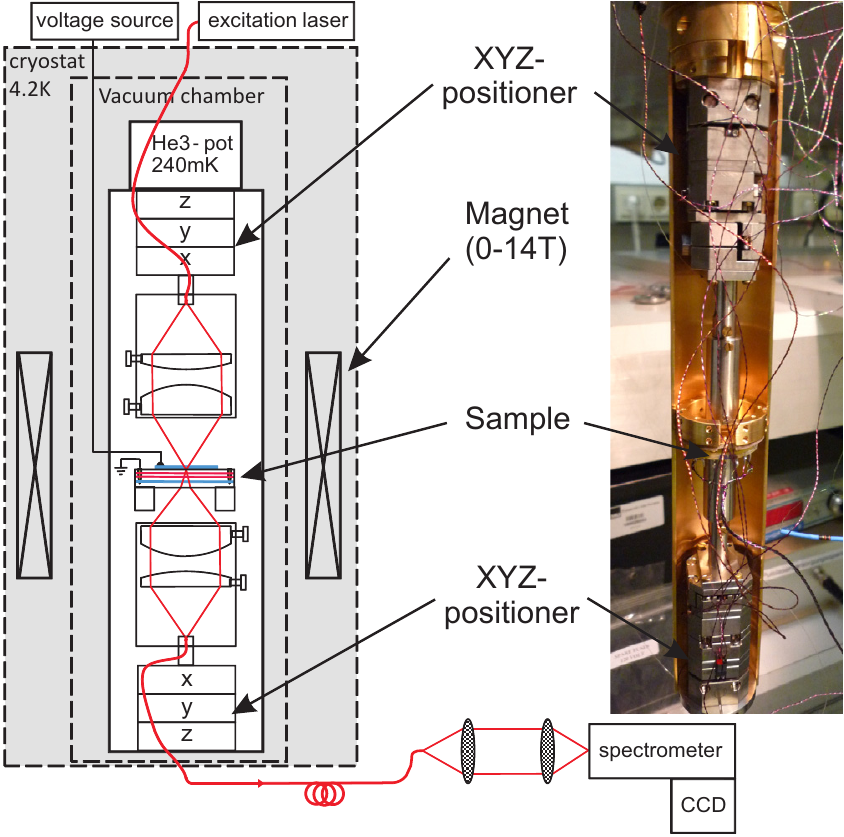}
\end{center}
\caption{\label{setup1} \textbf{Cryogenic setup.} Schematic diagram and photograph of our low temperature microscope with two diffraction limited confocal objectives embedded in a $^{\mathrm{3}}$He refrigerator.}
\end{figure}

In Fig.\ \ref{setup1} a scheme as well as a photograph of our experimental setup are displayed. The single mode fiber-based low-temperature microscope is composed of two confocal objectives. The microscope is embedded in a $^{\mathrm{3}}$He refrigerator with a base temperature below 240\,mK. The cryogenic setup exhibits an excellent long-time stability in position against temperature sweeps and magnetic field variations. Both objectives are diffraction limited and thus have a focus spot size with a diameter comparable to the wavelength and can be individually positioned with a low-temperature piezo-$xyz$-positioning unit \cite{2008Hoegele}. The positioning works at temperatures down to 240\,mK along all three space coordinates over millimeter distances and with a nanometer accuracy. The objective on top of the sample serves to generate excitons with laser radiation. The second objective below the sample is used to collect the photoluminescence (PL) light in transmission and analyze it outside the cryostat, e.g. in a grating spectrometer. In order to efficiently cool the device chip, it is mounted on a cold gold plate and bonded several times to a cold ground.

\section{Resonant exciton excitation}\vspace{2mm}

\begin{figure}[t]
\begin{center}
\includegraphics{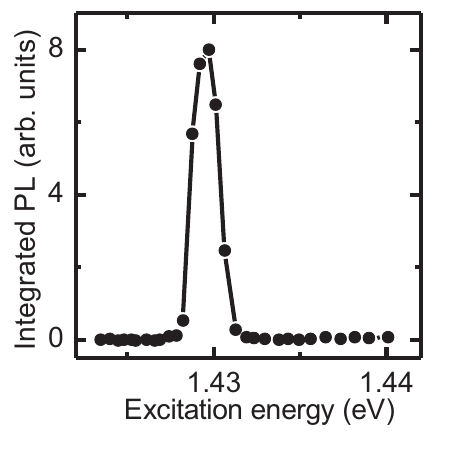}	
\end{center}
\caption{\textbf{Photoluminescence excitation measurement.} Integrated PL intensity centered at 1.399\,eV of the indirect excitons as a function of the laser excitation energy [$V_{\mathrm{Gate}}$ = 0.35\,V, $P_{\mathrm{Laser}}$($E_{\mathrm{Laser}}$) = 4.75\,$\mu$W, $T_{\mathrm{Lattice}}$ = 246\,mK].}
\label{PLE1}
\end{figure}

In our trapping configuration the exciton generation occurs on the slide gate typically spatially separated by 9\,$\mu$m from the center of the trap and the exciton trap is filled only with indirect excitons, pre-cooled to lattice temperature \cite{2011Schinner}. To further reduce the radiative heating in the sample and the excitation of free charge carriers, we resonantly pump the intra well direct exciton with a tunable diode laser. The photon energy of the excitation laser is below the GaAs band gap and, apart from some absorption in the semitransparent metal gates, can be only absorbed in the coupled double quantum wells. Accordingly, only the energy of the direct to indirect exciton conversion is deposited in the phonon bath of the sample. In Fig.\ \ref{PLE1} a photoluminescence excitation (PLE) measurement at $T_{\mathrm{Lattice}}$ = 246\,mK is shown. Here, the integrated PL intensity of the indirect excitons is plotted as a function of the laser excitation energy exciting direct excitons on an unstructured gate 10\,$\mu$m away from the detection spot. In Fig.\ \ref{PLE1} we observe a strong IX PL when the direct exciton is resonant with the tunable laser energy and so direct excitons are efficiently pumped transforming into the detected indirect excitons.

\section{Density dependence of the correlated excitonic photoluminescence}\vspace{2mm}
\begin{figure}[b]
\begin{center}
\includegraphics[width=8.6cm]{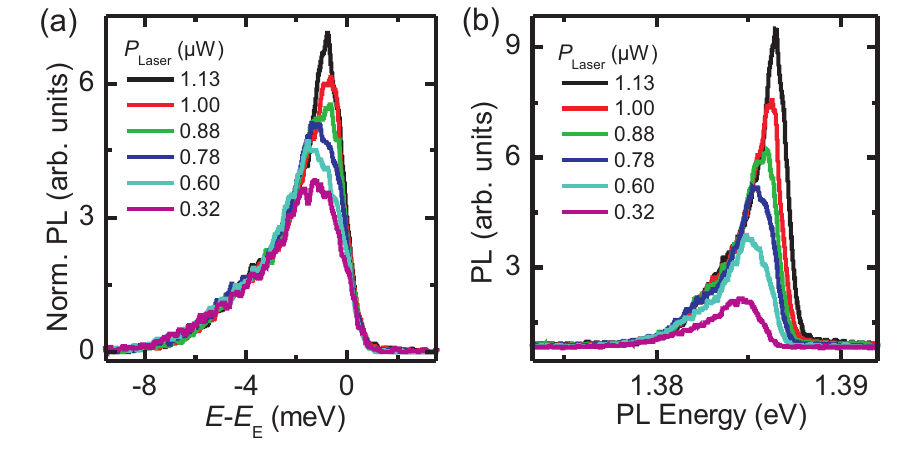}	
\end{center}
\caption{\textbf{Laser power dependence of the PL lineshape normalized versus original spectra.} (a) Energy resolved PL spectra for different laser excitation powers (identical to Fig.\ 3(a) in the main text). The PL intensity is normalized to the excitation laser power and the spectra are energetically shifted to correct the density caused blue shift. In (b) the corresponding original spectra are shown [$V_{\mathrm{G}}$ = 0.70\,V, $V_{\mathrm{T}}$ = 0.32\,V, and $V_{\mathrm{S}}$ = 0.35\,V, $T_{\mathrm{Lattice}}$ = 243\,mK, $E_{\mathrm{Laser}}$ = 1.4289\,eV].}
\label{spec}
\end{figure}

To explore the tunability of the exciton density in the trap with illumination intensity, we vary the incident laser power focused on the slide quasi-statically by tuning the power transmitted into the microscope with an acousto-optical modulator. Thus it is possible to change the excitation power by three orders of magnitude. Fig.\ \ref{spec}(a) is identical to Fig.\,3(a) in the main text and plots PL spectra, for different laser powers, detected from the center of the trap. The PL intensity is normalized to the excitation laser power and the spectra are energetically shifted such that the Fermi-edge-like steepest slopes in the PL vs. $E$ are $E_{\mathrm{E}} \approx 0$. The corresponding original spectra are shown in Fig.\ \ref{spec}(b). To obtain these spectra we integrate 55 seconds with a detection rate of 12.5 counts per second in the PL maximum at 1.0\,$\mu$W.

\section{Conversion from blue shift to exciton density}\vspace{2mm}

\begin{figure}[b]
\begin{center}
\includegraphics{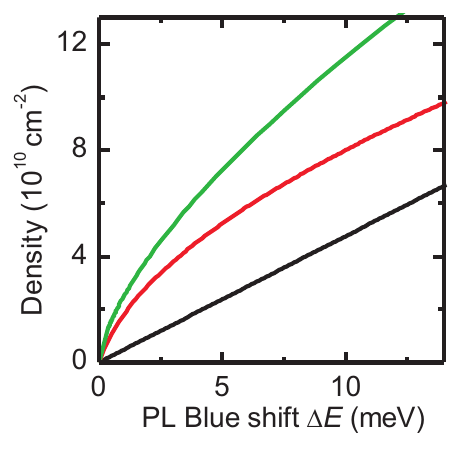}	
\end{center}
\caption{\textbf{Conversion from blue shift to exciton density.} The black straight line shows the linear conversion between the blue shift $\Delta E$ of the PL maximum energy and the corresponding IX density. In the red curve a density dependent correction factor is included providing a better approximation of the real exciton density. The upper green curve show the conversion between blue shift and trapped IX density for a two-dimensional hexagonal ordered crystal of parallel oriented dipolar excitons.}
\label{density1}
\end{figure}

The quantum confined Stark effect causes a red shift of the excitonic energy of $\Delta E_{\mathrm{IX}}=-\mathbf{pF}$ where $\mathbf{p}$ is the IX dipole moment and $\mathbf{F}$ the electric field perpendicular to the QW plane. In a trapped IX ensemble all individual dipoles $\mathbf{p}=e\mathbf{d}$ are aligned and perpendicular to the QW plane. The effective electric field, seen by the individual dipole moment of a single IX, is the sum of the externally applied electric field and the depolarizing dipole field of all other IX in the vicinity. With increasing exciton density the effective electric field decreases and causes a blue shift $\Delta E$. For not too small densities $n$ of $n\gtrsim 2.5\times10^{10}\frac{1}{\mathrm{cm}^{2}}$ the simplest approximation is that the blue shift is proportional to the exciton density $\Delta E = \frac{4\pi e^{2}d}{\epsilon}n$ (see black line in Fig.\ \ref{density1}) where $\epsilon$ is the dielectric constant. In the low-density limit the IX-IX interaction was theoretically investigated in Refs.\ \cite{2007Zimmermann, 2008Schindler, 2009Laikhtman, 2010Comment_Ivanov} and results in a density and temperature-dependent correction factor $f(T,n)$ in $\Delta E = \frac{4\pi e^{2}d}{\epsilon}nf(T,n)$. From this formula we find a nonlinear relation between blue shift and density as plotted by the red curve in Fig.\ \ref{density1}. The latter is used to convert blue shift $\Delta E$ to density in Fig. 3(b) of the main text.

A third naive $T=$ 0 approach to convert the measured blue shift into a trapped IX density is to sum up the dipole fields of the trapped indirect exciton ensemble acting on a single dipolar exciton. For this method it is assumed that the dipolar exciton liquid is ordered in a two-dimensional hexagonal crystal of dipoles arranged in parallel. A single dipolar exciton sees a blue shift, caused by another dipolar exciton in a distance $d_{\mathrm{IX-IX}}$, of
$W=\frac{p^2}{\epsilon d_{\mathrm{IX-IX}}^{3}}$ where $p=ed$ is the IX dipole moment resulting from the electron-hole separation by a distance $d$ in the two adjacent quantum wells and $\epsilon$ is the dielectric constant. The total blue shift $\Delta E$ for a single exciton with a next neighbor distance $d_{\mathrm{IX}}$ is calculated by summing up all contributions $W$ of every trapped exciton. The next neighbor distance $d_{\mathrm{IX}}$ is converted by $n= \frac{3\sqrt{3}}{4d_{\mathrm{IX}}^{2}}$ in the corresponding trapped indirect exciton density $n$. The green curve in Fig.\ \ref{density1} shows the result.

\section{Diamagnetic shift of the exciton energy in magnetic fields and relevant length scales}\vspace{2mm}
\begin{figure}[b]
\begin{center}
\includegraphics{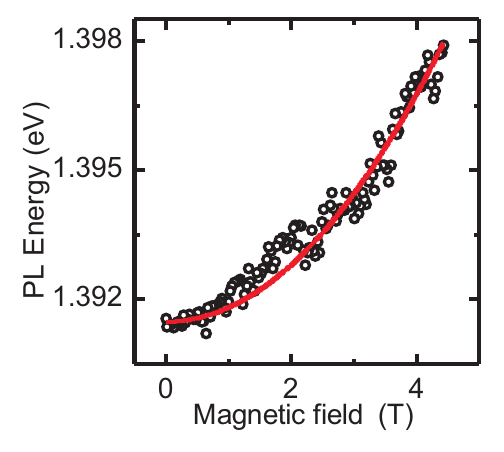}	
\end{center}
\caption{\textbf{Magnetic field dependence.} The PL maximum detected in the center of the trap is plotted as a function of the magnetic field applied perpendicular to the QW plane. In addition, a quadratic fit for the diamagnetic shift is shown from which we find an effective Bohr radius $r_{\mathrm{e}} =$ 20\,nm [$V_{\mathrm{G}}$ = 0.70\,V, $V_{\mathrm{T}}$ = 0.32\,V, and $V_{\mathrm{S}}$ = 0.35\,V, $T_{\mathrm{Lattice}}$ = 256\,mK, $E_{\mathrm{Laser}}$ = 1.4289\,eV, $P_{\mathrm{Laser}}$ = 3.5\,$\mu$W circular polarization].}
\label{dia}
\end{figure}

A magnetic field $B$ applied in Faraday configuration perpendicular to the QW plane causes a diamagnetic quadratical shift of the electrostatically trapped IX energy by $\Delta E _{\mathrm{dia}}= \frac{e^2 r_{\mathrm{e}}^2}{8 \mu c^2}B^2$ where $r_{\mathrm{e}}$ is the effective Bohr radius, $\mu$ the reduced mass of the exciton \cite{1986Bugajski, 2008SternMott}. In Fig.\ \ref{dia} the quadratic behavior of the PL maximum energy, detected from the center of the trap, is shown as a function of the magnetic field. Fitting the exciton diamagnetic shift we find a corrected effective Bohr radius of $r^{*}_{\mathrm{e}}=20$\,nm \cite{1998Walck}. If the hole is strongly localized, it is necessary to use the electron mass instead of the reduced mass $\mu$ which results in an effective Bohr radius of $r^{\mathrm{**}}_{\mathrm{e}}=16$\,nm. The quadratic diamagnetic shift shows that electrons and holes are excitonically bound. In our trapping configuration it is only possible to confine IX, free electrons are repelled. For free charge carriers or an electron-hole plasma different dependence of the PL line is reported, reflecting  quantized Landau levels \cite{2008SternMott, 2011Kowalik}.

The thermal de Broglie wavelength $\lambda_{\mathrm{dB}}$ of a bosonic particle in two dimensions with a quadratic dispersion relation is given by $\lambda_{\mathrm{dB}} = \frac{\hbar}{\sqrt{2m\pi kT}}$ where $m$ is the total mass of the exciton (sum of electron and hole mass) \cite{2000Yan}. The interexcitonic distance $d_{\mathrm{IX}}$ in a two-dimensional exciton lattice hexagonal arrangement is given by $d_{\mathrm{IX}}=\sqrt{\frac{3\sqrt{3}}{4n}}$. In the correlated many-body state the thermal de Broglie wavelength ($\approx 75$\,nm at 2\,K) is larger than the interexcitonic distance $d_{\mathrm{IX}} \approx 57$\,nm at $n=4\times10^{10}\frac{1}{\mathrm{cm}^{2}}$ and $d_{\mathrm{IX}}$ is similar to twice the effective Bohr radius $r_{\mathrm{e}} \approx$ 20\,nm.

\section{Analysis of the red tail of the spectra}\vspace{2mm}

The analysis of the red tail of the photoluminescence spectra clearly showed the occurrence of different power laws with distinct exponents. The first exponent with a value of about -0.1 can be related to a low-energy excitation of up to 1.5\,meV and reflects slow temporal behavior, possibly induced by thermal broadening. The second exponent shows values of about -0.8 describing a medium energy excitation of the IX ensemble in the energy interval between 1.5 and 4\,meV.
This power law rather abruptly disappears at $T_{\mathrm{Lattice}} \gtrsim$ 5\,K [Fig.\ 2(e) in the main text] and we relate it to a correlated many-body phase of a liquid of parallel oriented dipolar excitons as theoretically predicted in Ref.\ \cite{2009Laikhtman}.
This correlated phase appears only at low temperatures and high enough trapped IX densities. The third exponent of about -3 is related to high-energy ($E_{\mathrm{0}}-E > $ 4\,meV) excitations of the IX ensemble on a fast time scale. Most likely, phonon emission on a picosecond time scale and
shake-up processes, in which the recombining exciton transfers part of its energy to a collective excitation of its interacting neighboring excitons, are responsible for this behavior.

\begin{figure}[b]
\begin{center}
\includegraphics[width=8.6cm]{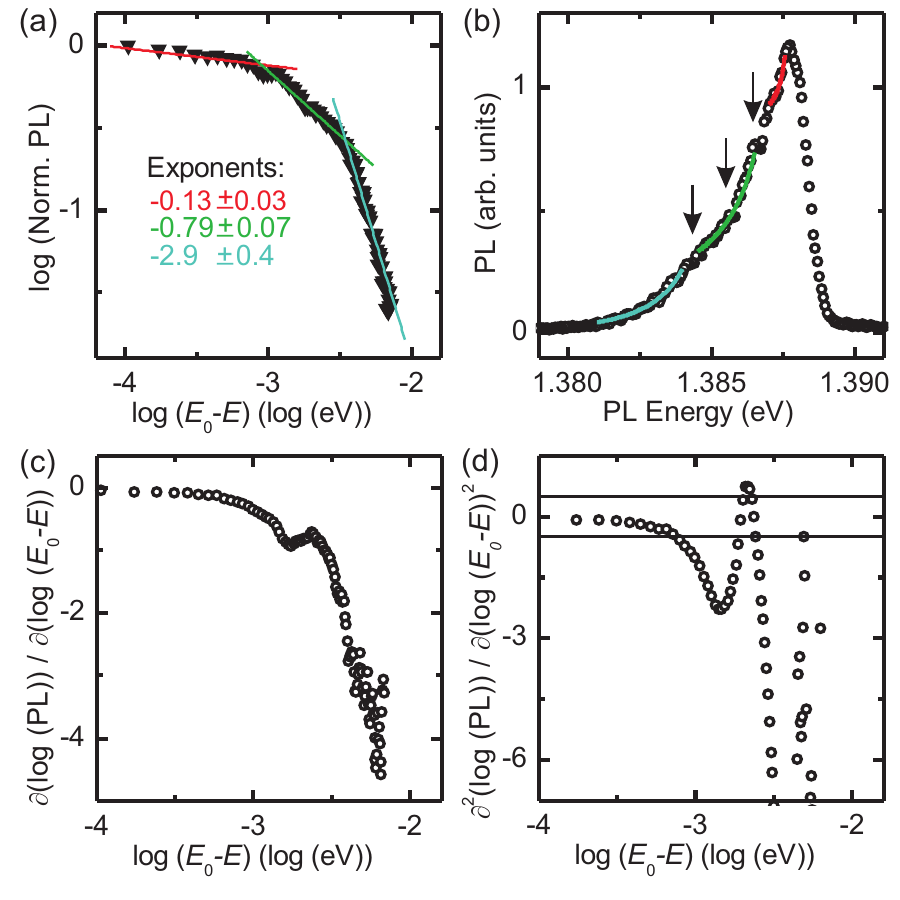}
\end{center}
\caption{\textbf{Analysis of the red tail of the spectra in a log-log diagram.} (a) $\log$-$\log$ plot of one typical spectrum. The slopes are determined by the numerical method described in the text [$V_{\mathrm{G}}$ = 0.70\,V, $V_{\mathrm{T}}$ = 0.32\,V, and $V_{\mathrm{S}}$ = 0.34\,V, $P_{\mathrm{Laser}}$ = 1.69\,$\mu$W, $T_{\mathrm{Lattice}}$ = 248\,mK, $i=5$]. (b) Original spectrum with power laws determined by the slopes. (c) First derivative of the logarithmized data.  (d) Second derivative of the logarithmized data. The values between the two horizontal lines represent data points which were used to calculate the slopes.}
\label{slopes}
\end{figure}

In order to determine the exponent of the different power laws $I(E) \sim (E_{0}-E)^{-|\alpha|}$ in the red tails of the PL lineshape in the log-log plots [Fig.\ \ref{slopes}(a), (b)] we employed the following numerical method. In Fig.\ \ref{slopes} this method is shown exemplarily for one spectrum. In the first step the first derivative of the logarithmized data is calculated. Simply taking the difference between two neighboring data points would yield a very noisy result. For a fixed energy $E_{\mathrm{0}}-E$ a linear fit is done over $\pm i$ neighboring data points. The slope obtained by the linear fit over the $2i+1$ data points gives the value of the first derivative for this energy. This is repeated successively for each energy point. The result is shown in Fig.\ \ref{slopes}(c). The second derivative is calculated in the same way from data points of the first derivative. All data pairs whose values of the second derivative are in a range of $\pm 0.5$ are considered as having a constant slope. These data points are between the two horizontal lines in Fig.\ \ref{slopes}(d). If the slopes of two or more data points differ less than 0.1 for the first slope and 0.5 for all other slopes these slopes and their corresponding energies are averaged. The previous values provide the best results but small variations do not change the exponents of the power laws. In contrast to Fig.\ 2(d) in the main text in Fig.\ \ref{slopes}(a) the logarithmized data are plotted to illustrate that in our numerical method a linear fit of the logarithmized data is done.

We used a second method as a consistency check by performing a linear fit over $\pm i$ neighboring data points. If none of the $2i+1$ residues is larger than a predefined value, these data points are considered as having a constant slope. This procedure is also repeated successively for each energy point, and data points with similar slopes are averaged in the same way as described above. Both methods yield comparable results. However, the first method turned out to be more general since the signal-to-noise ratios of our spectra differ substantially and the residues depend strongly on the signal-to-noise ratio.

\section{Interference substructure superimposed on the PL lineshape}\vspace{2mm}

Photoluminescence spectra having many photon counts and correspondingly large signal-to-noise ratio show a pronounced substructure in the form of energetically equidistant shoulders. In Fig.\ \ref{slopes}(b) this shoulders are marked with arrows. While the spectra shift with changing IX density the substructure is fixed on an absolute energy scale. We identify the energetic periodicity of the substructure as an interference effect inside the heterostructure, where a cavity is formed between the top and the bottom surface of the sample. A similar interference effect has been reported in \cite{2011Latta}. We found that the slopes are influenced by the substructure and the slope starts to oscillate. Nevertheless, the data at low densities yield a convincing evidence that the power law with -$|\alpha| \approx$ -0.8 is an intrinsic and characteristic feature and is not caused by such interference effects. All temperature-dependent measurements show that the slope slightly lower than 1 disappears for temperatures $T\gtrsim5$\,K. This suggests a breakdown of the correlation effect and cannot be explained with an interference effect. In log-log-plots of spectra with normal signal-to-noise ratios we observe no pinning of the boundary point positions on the interference substructure. We explain this with the absence of the substructure due to the lower signal-to-noise ratio.

\vspace{3mm}

\section{Indirect exciton gas temperature}\vspace{2mm}
\begin{figure}[b]
\begin{center}
\includegraphics[width=8.6cm]{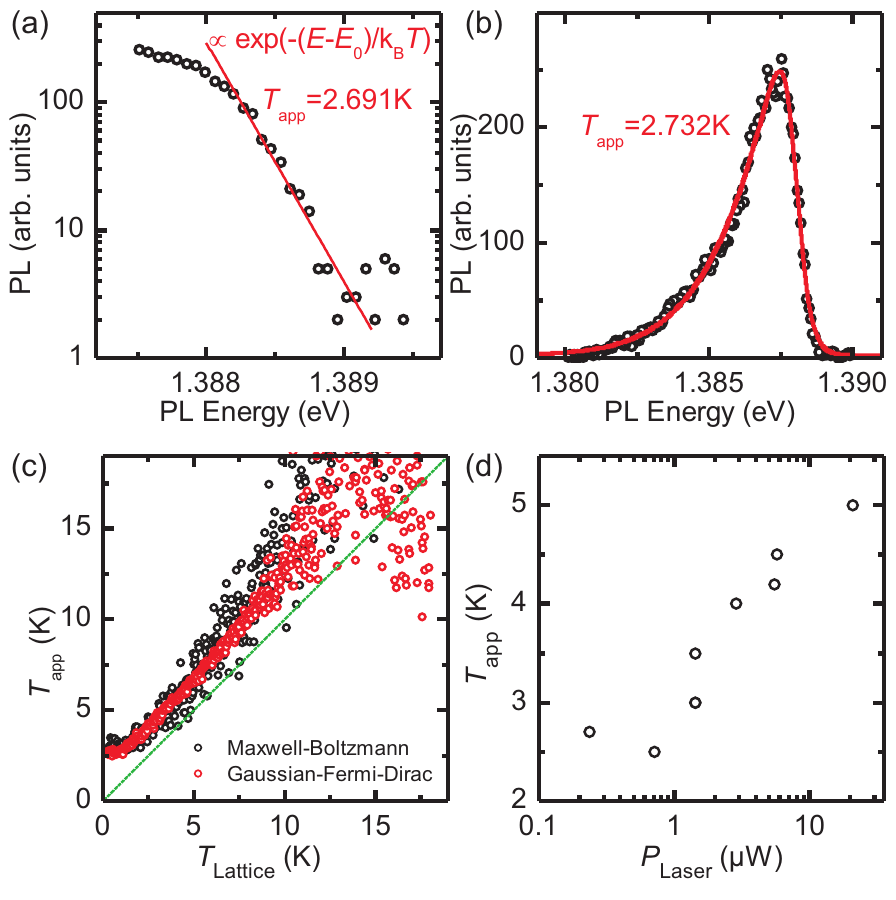}	
\end{center}
\caption{\textbf{Extraction of an apparent exciton gas temperature.} (a) Natural logarithm plot of the blue side of one spectrum. The red line is obtained by the numerical method described in the text [$V_{\mathrm{G}}$ = 0.70\,V, $V_{\mathrm{T}}$ = 0.32\,V, and $V_{\mathrm{S}}$ = 0.35\,V, $P_{\mathrm{Laser}}$ = 1.4\,$\mu$W, $T_{\mathrm{Lattice}}$ = 285\,mK, $i = 5$]. (b) PL spectrum from the same data as in (a). The red curve represents a \textquotedblleft Gauss-Fermi-Fit\textquotedblright, which is described in the text. (c) Fit results of a temperature dependent measurement. Both methods are in good agreement. (d) Saturation temperatures of $T_{\mathrm{app}}$ for different excitation powers extracted from different temperature sweeps like those displayed in (c) [$V_{\mathrm{G}}$ = 0.70\,V, $V_{\mathrm{T}}$ = 0.32\,V, and $V_{\mathrm{S}}$ = 0.35\,V].}
\label{temperature}
\end{figure}
We use two different methods to determine the apparent exciton gas temperature $T_{\mathrm{app}}$ [see also Fig.\,2(c) and Fig.\,3(c) in the main text] from the blue tail of the PL spectrum. In the first method the exciton gas temperature is extracted from the exponential Boltzmann-like behavior of the blue tail. Therefore the natural logarithm of the PL intensity is calculated and the slope of the linear regime is determined with the same numerical method as described above. One slope obtained with this method is shown in Fig.\ \ref{temperature}(a) together with the logarithmically plotted spectrum. Using logarithmic data for the fit has the advantage that data with low intensity away from the PL maximum are weighted more strongly compared to an exponential regression of the original data. In this regime the assumption of a Maxwell-Boltzmann distribution is approximately valid. The exciton gas temperature can be calculated directly from this slope. Our second approach is to multiply a Gaussian distribution with a Fermi-Dirac distribution. The Gaussian distribution is introduced to reflect broadening of the PL line by (i) the energy dependent density of states of the interacting IX dipoles in the trap (ii) the random disorder potential in the CDQW and (iii) effects of time-dependent spectral diffusion as the displayed spectra are accumulated over approximately 1\,minute. The Fermi-Dirac distribution is used as a cutoff function, reflecting the thermal smearing of the filling of the trap with the dipolar exciton liquid. In combination the Gaussian distribution describes the red tail of the spectrum while the Fermi-Dirac distribution describes the temperature-broadened steep decrease on the blue tail of the spectrum. In the case of the bosonic dipolar liquid the strong spatial dipolar repulsion replaces the role of the Pauli principle in the Fermi-Dirac distribution.
One spectrum together with its fit is shown in Fig.\ \ref{temperature}(b). Although the assumption of a Fermi sea can only be considered approximatively since we have an IX ensemble with hybridized electrons and no free electrons, the results of this method are in good agreement with the exciton gas temperatures obtained by the first method. The temperatures extracted by the Fermi-Dirac distribution scatter less than these using the Maxwell-Boltzmann fit. Both fits show a constant offset to the lattice temperature [green line in Fig.\ \ref{temperature}(c)] which at least in part can result from the broadening contributions discussed above which persist to lower lattice temperatures. Additional temperature independent broadening of the blue tail of the excitonic emission, not included in the modeling, is likely to result from scattering events or shake-up processes and may be an essential contribution to this constant offset of the exciton gas temperature extracted from the fits. Hence $T_{\mathrm{app}}$ should be considered as an upper limit of the IX temperature whereas the real IX temperature is likely to be considerably closer to the lattice temperature.
For a lattice temperature $T_{\mathrm{Lattice}} \lesssim$ 1 to 2\,K, we observe a saturation of the $T_{\mathrm{app}}$ [Fig.\ \ref{temperature}(c)] corresponding to an upper bound of the apparent exciton gas temperature at $T_{\mathrm{app}} = 3$\,K.
Fig.\ \ref{temperature}(d) shows the saturation temperatures of several temperature sweeps with different excitation powers. This plot indicates that the saturation temperature decreases exponentially with decreasing excitation power.

\end{document}